\documentclass[11pt]{article}

\usepackage{jheppub_mod}
\usepackage[english]{babel}
\usepackage{amsmath}
\usepackage{amssymb}
\usepackage{amsthm,amsfonts}
\usepackage{dsfont}
\usepackage{tensor}
\usepackage[dvipsnames]{xcolor}
\usepackage[mathstyleoff]{breqn}
\usepackage{slashed}
\usepackage{stmaryrd}
\usepackage{enumerate}
\usepackage[normalem]{ulem}
\usepackage{tikz}
\usepackage[mathscr]{euscript}
\usepackage{mathrsfs}
\usepackage{hyperref}
\usepackage{amsthm}
\usepackage{svrsymbols}
\usepackage{phaistos}
\usepackage{mathtools}
\usepackage{stmaryrd} 
\usepackage{graphicx}

\DeclareMathAlphabet{\mathpzc}{OT1}{pzc}{m}{it}
\def\ov{\overline}
\def\un{\underline}

\def\wt{\widetilde}
\def\be{\begin{equation}}
\def\ee{\end{equation}}
\def\bea{\begin{eqnarray}}
\def\eea{\end{eqnarray}}
\def\ap{\alpha'}
\newcommand{\nn}{\nonumber}

\title{Supersymmetric $\alpha'$-corrections to the Generalized Kerr–Schild Ansatz}

\author{Jesús A. Rodríguez ${}^{a,b}$}
\emailAdd{jarodriguez@df.uba.ar}
\affiliation{${}^{a}$ Universidad de Buenos Aires, Departamento de Física.\\${}^{} \ $ Ciudad Universitaria, 1428. Buenos Aires, Argentina.}
\affiliation{${}^{b}$ Universidad Argentina de la Empresa (UADE). Facultad de Ingeniería y Ciencias Exactas.\\${}^{} \ $ Departamento de Ciencias Básicas. Buenos Aires, Argentina.}

\abstract{We investigate the supersymmetric extension of the generalized Kerr-Schild ansatz (gKSA) in Double Field Theory (DFT) including first-order $\alpha'$ corrections. Supersymmetry plays a central role in constraining higher-derivative deformations, and in this work we focus on the structure of the $\mathcal{O}(\alpha')$ Killing Spinor Equations (KSEs). Starting from the $\alpha'$-corrected supersymmetry transformations of the fermionic fields, we derive the corresponding KSEs in DFT and analyze their behavior under the gKSA. The generalized Green–Schwarz transformations impose specific constraints on the perturbations that ensure the linearization of the $\alpha'$-corrected KSEs. We then provide the explicit parameterization of the results in terms of the ten-dimensional $\mathcal{N}=1$ supergravity fields. Our construction establishes a unified framework for exploring supersymmetric backgrounds in $\alpha'$-corrected DFT and offers new tools for generating higher-derivative supergravity solutions.
}

\begin{document}
\maketitle

\section{Introduction}

String theory exhibits a rich web of dualities that often remain hidden at the level of effective field theory. Among them, T-duality plays a central role, motivating the search for formulations in which this symmetry becomes manifest. Double Field Theory (DFT) \cite{Siegel:1993xq, Siegel:1993th, Hull:2009mi, Hohm:2010pp} achieves this by doubling the spacetime coordinates and adopting an $O(D,D)$-covariant framework. In doing so, it provides a unified geometric language that illuminates the structure of string theory, enables the study of non-geometric backgrounds, and deepens our understanding of dualities and their implications for effective field theories.

Significant progress has been made in extending the framework of DFT, including developments in dimensional reductions \cite{Hohm:2013nja, Betzler:2017kme}, gauged supergravities \cite{Grana:2012rr, Aldazabal:2013mya, Blumenhagen:2015lta, Lescano:2021but, Hatsuda:2022zpi}, and the study of non-geometric fluxes \cite{Andriot:2012wx, Dibitetto:2012rk, Andriot:2012an}. Further insights have emerged from worldsheet approaches \cite{Copland:2011wx, Driezen:2016tnz, Demulder:2018lmj}, supersymmetry and superspace techniques \cite{Hohm:2011nu, Jeon:2011sq, Berman:2013cli, Butter:2021dtu, Butter:2022gbc, Butter:2023nxm}, and the inclusion of higher-derivative corrections \cite{Hohm:2013jaa, Bedoya:2014pma, Hohm:2014xsa, Marques:2015vua, Baron:2017dvb, Baron:2018lve, Baron:2020xel, Lescano:2021guc}, among other directions. 

Understanding higher-derivative terms is particularly important for evaluating stringy effects in various gravitational scenarios. A systematic method to construct the infinite tower of $\alpha'$-corrections to the string effective action, known as the generalized Bergshoeff-de Roo (gBdR) identification, was developed in \cite{Baron:2018lve, Baron:2020xel}. While this method provides a powerful approach to generating correction terms related to the generalized Green-Schwarz transformation \cite{Marques:2015vua}, obstructions arise in fully determining the ${\cal O}(\alpha'{}^{3})$ contributions \cite{Hronek:2020xxi, Hsia:2024kpi}. Nevertheless, the gBdR identification has proven extremely useful in deriving the first two orders in the $\alpha'$ expansion, corresponding to four- and six-derivative corrections. Despite the challenges in formulating a fully $O(D,D)$-covariant description of higher-derivative corrections, DFT remains a valuable framework for exploring physics at leading  $\alpha'$ orders. In this context, searching for solutions in diverse scenarios such as cosmology \cite{Wu:2013sha, Wu:2013ixa, Ma:2014ala, Brandenberger:2017umf, Brandenberger:2018bdc, Bernardo:2019pnq, Angus:2019bqs, Lescano:2020smc, Lescano:2021nju, Angus:2024owg} and black hole physics \cite{Ko:2016dxa, Angus:2018mep, Edelstein:2019wzg, Liu:2021xfs, Li:2024ijj, Ying:2024jjr, Lunin:2024vsx} becomes a compelling objective. 

A powerful technique for simplifying and solving Einstein's equations in general relativity is the Kerr-Schild ansatz \cite{Kerr:1963ud, Debney:1969zz}. Its extension to DFT was introduced in \cite{Lee:2018gxc} and further developed in a series of works \cite{Cho:2019ype, Kim:2019jwm, Lescano:2020nve, Berman:2020xvs, Lescano:2021ooe, Angus:2021zhy, Lescano:2022nhp, Berman:2022bgy}. The core idea behind the (generalized) Kerr-Schild ansatz is to perturb a background metric, which is itself a solution of the theory, using null geodesic vectors, thus reducing the equations of motion to linear equations. In general relativity, the metric perturbation is linear and exact, along with its inverse. In the supergravity context, the generalized Kerr-Schild ansatz (gKSA) enables perturbations of the full NS-NS field content, albeit at the cost of introducing nonlinearity in the metric and the Kalb-Ramond field, while still preserving the linearization of the equations of motion \cite{Lee:2018gxc}. Moreover, this ansatz can be applied to perturb the fermionic fields in supersymmetric DFT \cite{Lescano:2020nve} and to incorporate higher-derivative corrections \cite{Lescano:2021ooe}. 

The ${\cal O}(\alpha')$ equations of motion obtained in \cite{Lescano:2021ooe} are highly intricate, making it challenging to obtain or even evaluate solutions. In the present work, we take an alternative approach to fully explore the potential of the gKSA for generating supergravity solutions. Beyond the equations of motion, the Killing spinor equations (KSEs) provide a valuable tool for identifying supersymmetric configurations. By reducing the problem to a first-order system, they often simplifies the search for solutions, especially when combined with a suitable ansatz for the (bosonic) fields. The effects of the gKSA on the KSEs at leading order in $\alpha'$ were studied in \cite{Lee:2018gxc}. Here, we extend this analysis to first order in $\alpha'$ by employing the supersymmetry transformations of the generalized fermionic fields computed in \cite{Lescano:2021guc} together with the results obtained from the generalized Green-Schwarz transformation.

This paper is organized as follows. The introduction is complemented by a brief review of the basic features of Double Field Theory, which also establishes the notation and conventions used throughout the work. Section 2 presents the Kerr–Schild ansatz in both general relativity and DFT, highlighting their similarities and differences. Section 3 addresses the $\alpha'$-corrections to DFT under the action of the gKSA and shows how the perturbation of the generalized Green–Schwarz transformation imposes key conditions on the dynamics of the theory. Section 4 discusses ${\cal N}=1$ supersymmetry in DFT, focusing on the Killing spinor equations under the gKSA and presenting our main result: the $\alpha'$-corrections to the KSEs. Section 5 provides the parameterization in $D=10$ in terms of ${\cal N}=1$ supergravity multiplets. We conclude in Section~6 with a discussion of our results and outlook.

\subsection{Basics of Double Field Theory}

Double Field Theory (DFT) is formulated on a space with coordinates $X^{M}$, which belong to the fundamental representation of the symmetry group $O(D,D)$. The indices $M,N,\dots=0,1,\dots 2D-1$, label the doubled space, where $D$ is the spacetime dimension. The theory is constructed using multiplets of the duality group, making $O(D,D)$ a manifest global symmetry. The invariant group metric $\eta_{MN}$
\be
\eta_{MN} = \left(\begin{matrix} 0 & \delta_{\mu}{}^{\nu}\\ 
\delta^{\mu}{}_{\nu} & 0 \end{matrix}\right)\, , \qquad \mu,\nu,\dots = 0,1,\dots D-1\, ,
\ee
and its inverse $\eta^{MN}$ are used to raise and lower indices in the doubled space. 

To ensure consistency, the theory imposes the strong constraint
\be
\label{dft-strong-constraint}
\partial_{M}\partial^{M}(\cdots) \ = \ \partial_{M}(\cdots)\partial^{M}(\cdots) \ = \ 0 \, ,
\ee
where $(\cdots)$ represents generalized fields or their combinations. This constraint restricts coordinate dependence, effectively removing half of the generalized coordinates and giving rise to $D$-dimensional theories.

The fundamental fields of DFT include the generalized metric $H_{MN}$, a symmetric tensor, and the generalized dilaton $d$, which together describe the massless bosonic sector of string theory. The generalized metric is parameterized as
\be
\label{dft-generalized-metric}
H_{MN} = \begin{pmatrix} g^{\mu \nu} & - g^{\mu \rho} b_{\rho \nu} \\ b_{\mu \rho} g^{\rho \nu} & g_{\mu \nu} - b_{\mu \rho} g^{\rho \sigma} b_{\sigma \nu}\end{pmatrix}\, ,
\ee
where $g_{\mu\nu}$ is the $D$-dimensional metric, and $b_{\mu\nu}$ is the Kalb-Ramond field. The generalized dilaton is defined as
\be
\label{dft-generalized-dilaton}
e^{-2d} = \sqrt{-g}e^{-2\phi}\, ,
\ee
where $g$ is the determinant of the metric and $\phi$ is the dilaton field.

Using the $O(D,D)$-invariant metric and the generalized metric, we define the projectors
\be
\label{dft-projectors}
P_{MN} = \frac{1}{2}\left(\eta_{MN} - H_{MN}\right)\, , \qquad \ov{P}_{MN} = \frac{1}{2}\left(\eta_{MN} + {H}_{MN}\right)\ ,
\ee
which allow decomposition of vectors as $V_{M}=\ov{P}_{M}{}^{N}V_{N}+P_{M}{}^{N}V_{N}=V_{\ov{M}} + V_{\un{M}}$.

\subsubsection*{Frame formulation}

To introduce fermionic fields, local Lorentz symmetries must be incorporated. In DFT, this is achieved through the generalized frame formalism, which enables the inclusion of fermions and facilitates the computation of higher-order derivative corrections.

In this formulation, the fundamental field is the generalized frame $E_{M}{}^{A}$, which parameterizes the coset $O(D,D)/O(D-1,1)_{L}\times O(1,D-1)_{R}$, where $O(D-1,1)_{L}\times O(1,D-1)_{R}$ accounts for the duplication of local Lorentz symmetries needed to describe the geometric properties of the doubled space. The generalized frame satisfies the constraints
\be
\label{dft-frame-constraints}
\eta_{MN} = E_{M}{}^{A}\eta_{AB}E_{N}{}^{B}, , \qquad H_{MN} = E_{M}{}^{A}H_{AB}E_{N}{}^{B}\, ,
\ee
where the flat indices $A$ decompose as $A=\left(\un{a},\ov{a}\right)$, with $\un{a},\ov{a}=0,1,\dots,D-1$. The metrics $\eta_{AB}$ and $H_{AB}$ 
\be
\label{dft-h-metrics}
\eta_{AB} = \left(\begin{matrix} -g_{\un{ab}} & 0\\
0 & g_{\ov{ab}}\end{matrix}\right)\, , \quad H_{AB} = \left(\begin{matrix} g_{\un{ab}} & 0\\
0 & g_{\ov{ab}}\end{matrix}\right)\, , \quad g_{\un{ab}} = g_{\ov{ab}} = \text{diag}(-1, \underbrace{+1, \dots, +1}_{D-1 \text{ times}})\, ,
\ee
are constant, invertible, and invariant under the double Lorentz group. The metric $\eta_{AB}$ is used to raise and lower flat indices. The parameterization of the generalized frame is given by
\be
\label{dft-generalized_frame}
E_{M}{}^{A} = \frac{1}{\sqrt{2}}\left(\begin{matrix} e^{\mu}{}_{\un{a}} & e^\mu{}_{\ov{a}} \\
- e_{\mu\un{a}} + b_{\mu\nu}e^{\nu}{}_{\un{a}} \  &  \ e_{\mu\ov{a}} + b_{\mu\nu}e^{\nu}{}_{\ov{a}}\end{matrix}\right)\, ,
\ee
where the $D$-dimensional vielbeins $e_{\mu}{}^{\un{a}}$ and $e_{\mu}{}^{\ov{a}}$ satisfy
\be
\label{dft-nd_vielbein}
e_{\mu}{}^{\un{a}}g_{\un{ab}}e_{\nu}{}^{\un{b}} = e_{\mu}{}^{\ov{a}}g_{\ov{ab}}e_{\nu}{}^{\ov{b}} = g_{\mu\nu}\,.
\ee
which will be identified as $e_{\mu}{}^{\un{a}}\delta_{\un{a}}^{a}=e_{\mu}{}^{\ov{a}}\delta_{\ov{a}}^{a}=e_{\mu}{}^{a}$ after fixing the gauge, where $e_{\mu}{}^{a}$ is the supergravity vielbein, and the indices $a,b,\dots=0,\dots,D-1$ are indices of $O(1,D-1)$, thus reducing the double Lorentz group to the conventional Lorentz group.

As usual, the incorporation of local symmetries requires considering covariant derivatives. Thus, due to the double Lorentz symmetry, the covariant derivative is defined as
\be
\label{dft-covariant}
\nabla_{M}V_{A} = \partial_{M}V_{A} + \omega_{MA}{}^{B}V_{B}\, ,
\ee
with $\omega_{MA}{}^{B}$ the generalized spin connection, satisfying
\be
\omega_{MAB} = -\omega_{MBA}\, , \qquad \omega_{M\un{a}\ov{b}} = \omega_{M\ov{a}\un{b}} = 0\, .
\ee
Unlike in standard Riemannian geometry, not all components of the generalized spin connection in DFT are determined. However, requiring the covariant derivative to be compatible with the generalized frame, $\nabla_{M}E_{N}{}^{A}=0$, allows us to define the generalized fluxes in terms of fully determined components of the spin connection
\be
\label{dft-generalized-fluxes}
F_{ABC} = -3\omega_{[ABC]} = 3D_{[A}E^{M}{}_{B}E_{|M|C]}\, , \qquad F_{A} = -\omega_{BA}{}^{A} = \sqrt{2}e^{2d}\partial_{M}\left(E^{M}{}_{A}e^{-2d}\right)\, ,
\ee
where $D_{A}=\sqrt{2}E^{M}{}_{A}\partial_{M}$ is the flat derivative. These fluxes play a key role in the dynamics of the theory, which is governed by the action
\be
\label{dft-action}
S = \int d^{2D}X \ e^{-2d}{\cal{R}}\, ,
\ee
where the generalized Ricci scalar $\mathcal{R}$ can be expressed in terms of the fluxes as
\be
\label{dft-generalized_ricci}
{\cal{R}} = 2D_{\un{a}}F^{\un{a}} + F_{\un{a}}F^{\un{a}} - \frac{1}{6}F_{\un{abc}}F^{\un{abc}} - \frac{1}{2}F_{\ov{a}\un{bc}}F^{\ov{a}\un{bc}}\, .
\ee

Now, let us consider the equations of motion. Recall that Double Field Theory is a constrained theory, and variations of the generalized frame are not entirely arbitrary but are subject to constraints. Taking this into account, the variation of the action with respect to the generalized dilaton and the generalized frame leads to the following equations of motion
\be
\label{dft-EOM's-frame}
{\cal R} = 0\, , \qquad {\cal R}_{\ov{a}\un{b}} = 0\, ,
\ee
where ${\cal R}$ is given by \eqref{dft-generalized_ricci} and 
\be
\label{dft-ricci-scalar-flat}
{\cal R}_{\ov{a}\un{b}} = D_{\ov{a}}F_{\un{b}} - D_{\un{c}}F_{\ov{a}\un{b}}{}^{\un{c}} + F_{\un{c}\ov{d}\ov{a}}F_{\un{b}}{}^{\un{c}\ov{d}} - F_{\un{c}}F_{\ov{a}\un{b}}{}^{\un{c}}\, .
\ee 

\subsubsection*{${\cal N}=1$ supersymmetry}

Double Field Theory naturally admits supersymmetric extensions \cite{Hohm:2011nu, Jeon:2011sq}. In the case of ${\cal N}=1$, $D=10$, the field content includes two Majorana-Weyl fermions: the generalized gravitino $\Psi_{\ov{a}}$, which transforms as a Majorana-Weyl spinor under ${\rm{O}}(9,1)_{L}$ and as a vector under $\rm{O}(1,9)_{R}$; and the generalized dilatino $\varrho$, which is a Majorana-Weyl spinor under ${\rm{O}}(9,1)_{L}$ and a scalar under $\rm{O}(1,9)_{R}$. Both fields are scalars under the transformations of the ${\rm{O}}(10,10)$ group. For an arbitrary $O(9,1)_{L}$ generalized spinor, the covariant derivative acts as
\be
\label{dft-covariant-spinor}
\nabla_{A}\Sigma = D_{A}\Sigma - \frac{1}{4}\omega_{A\un{bc}}\gamma^{\un{bc}}\Sigma\, ,
\ee
where $\gamma^{\un{ab\cdots}}=\gamma^{[\un{a}}\gamma^{\un{b}}\cdots{\gamma}^{\cdots]}$ denotes the antisymmetric product of generalized gamma matrices. These matrices satisfy the Clifford algebra of $O(9,1)_{L}$
\be
\label{clifford-algebra}
\left\{\gamma^{\un{a}},\gamma^{\un{b}}\right\} = - 2P^{\un{ab}}\, .
\ee

In addition to the usual symmetries of the theory, we now have supersymmetry transformations. These transformations are parametrized by a Majorana-Weyl spinor $\epsilon$, and are given by
\begin{align}
\label{dft-susy-transformations}
\delta_{\epsilon}E_{M}{}^{A} & = -\ov{\epsilon}\gamma^{[A}\Psi^{B]}E_{MB}\, , & \delta_{\epsilon}d & = - \frac{1}{4}\ov{\epsilon}\rho\, , \\
\delta_{\epsilon}\Psi_{\ov{a}} & = \nabla_{\ov{a}}\epsilon\, , & \delta_{\epsilon}\varrho & = - \gamma^{\un{a}}\nabla_{\un{a}}\epsilon\, \nn ,
\end{align}
which reflect how the bosonic and fermionic fields couple to each other in the framework of ${\cal N}=1$ supersymmetry.

With the field content and transformation rules established, we can now construct the action for supersymmetric DFT. To the lowest order in fermions, the action takes the form
\be
\label{dft-N=1_DFT_action}
S = \int \ d^{20}X \ e^{-2d}\left({\cal{R}} + \ov{\Psi}^{\ov{a}}\gamma^{\un{b}}\nabla_{\un{b}}\Psi_{\ov{a}} - \ov{\varrho}\gamma^{\un{a}}\nabla_{\un{a}}\varrho + 2\ov{\Psi}^{\ov{a}}\nabla_{\ov{a}}\varrho\right)\, ,
\ee
where ${\cal R}$ is the generalized Ricci scalar given by \eqref{dft-generalized_ricci}, and the fermionic terms represent the kinetic and interaction contributions of the fermionic fields.

\section{The Kerr-Schild ansatz}

In this section, we introduce some important aspects concerning the Kerr-Schild ansatz. We begin by reviewing its formulation in general relativity and then extend the discussion to Double Field Theory, highlighting key differences and novel features that arise in the duality-covariant framework.

\subsection{The Kerr-Schild ansatz in General Relativity}

The Kerr-Schild ansatz expresses the $D$-dimensional spacetime metric $g_{\mu\nu}$ as a perturbation of a background metric $\widetilde{g}_{\mu\nu}$, which satisfies the vacuum Einstein equations but is not necessarily the Minkowski metric
\be
\label{gr-ks-ansatz}
g_{\mu\nu} = \widetilde{g}_{\mu\nu} + \kappa\varphi l_{\mu}l_{\nu}\, .
\ee
Here, $\kappa$ is an expansion parameter, $\varphi$ is a scalar function depending on the coordinates and $l_{\mu}$ is a null vector\footnote{A null vector has no magnitude but can be given a direction. A null vector is orthogonal to itself because $l\cdot l = \wt{g}^{\mu\nu}l_{\mu}l_{\nu} = 0$} 
\be
\label{gr-null-condition}
g^{\mu\nu}l_{\mu}l_{\nu} = \widetilde{g}^{\mu\nu}l_{\mu}l_{\nu} = 0\, .
\ee
with respect to both $g$ and $\widetilde{g}$. The nullity of $l_{\mu}$ ensures that the perturbation term is highly constrained, leading to significant simplifications.

An important consequence is the simple expression for the inverse metric. To compute it, we start with
\be
g^{\mu\nu} = 
\left(\widetilde{g}_{\mu\nu} + \kappa\varphi l_{\mu}l_{\nu}\right)^{-1} = \left(\delta^{\rho}{}_{\nu} + \kappa\varphi l^{\rho}l_{\nu}\right)^{-1}\widetilde{g}^{\mu\rho}\, . \nn
\ee
Expanding the inverse term as a Taylor series gives
\be
\left(\delta^{\rho}{}_{\nu} + \kappa\varphi l^{\rho}l_{\nu}\right)^{-1} = \left(\delta_{\rho}{}^{\nu} - \kappa\varphi l_{\rho}l^{\nu} + \kappa^{2}\varphi^{2}l_{\rho}l^{\lambda}l_{\lambda}l^{\nu} + \cdots\right)\, , \nn
\ee
and hence, the nullity of $l_{\mu}$ indicates that all the terms of $\mathcal{O}(\kappa^{2})$ and higher vanish. This leads to the compact expression
\be
\label{gr-ks-inverse}
g^{\mu\nu} = \widetilde{g}^{\mu\nu} - \kappa\varphi l^{\mu}l^{\nu}\, .
\ee
Thus, the inverse metric retains a simple structure, closely related to the background metric. For consistency, indices of $l_{\mu}$ are raised and lowered using either $g$ or $\wt{g}$
\be
\label{gr-metric-determinant}
l_{\mu} = g_{\mu\nu}l^{\nu} = \widetilde{g}_{\mu\nu}l^{\nu}\, , \qquad l^{\mu} = g^{\mu\nu}l_{\nu} = \widetilde{g}^{\mu\nu}l_{\nu}\, . \nn
\ee

Another remarkable property of the Kerr-Schild ansatz is that the determinant of the metric remains unchanged
\be
\mathrm{det}\left(g\right) = \mathrm{det}\left(\widetilde{g}\right)\, .
\ee
This property simplifies calculations involving volume forms and curvature invariants.

One of the key advantages of the Kerr-Schild ansatz is its ability to linearize Einstein's equations. This is a consequence of a consistence condition on the null vector, which must also satisfy the geodesic equation for both metrics
\be
\label{gr-geodesic}
l^{\mu}\wt{\nabla}_{\mu}l_{\nu} = 0\, , \qquad l^{\mu}\nabla_{\mu}l_{\nu} = 0\, .
\ee
These geodesic conditions ensures that the perturbation does not introduce additional nonlinearities, making it easier to solve the Einstein equations. So, considering both, the null and geodesic conditions on $l_{\mu}$, the vacuum Einstein equation reads
\be
\label{gr-einstein-vacuum}
R_{\mu\nu} = \kappa R^{(1)}_{\mu\nu} + \kappa^{2}\varphi l_{\mu}l^{\rho}R^{(1)}_{\rho\nu} = 0 \, ,
\ee
where $R^{(1)}$ denotes the linear terms in $\kappa$,
\be
\label{gr-R1-linear}
R^{(1)}_{\mu\nu} = \frac{1}{2}\kappa\wt{\nabla}^{\rho}\left(\wt{\nabla}_{\mu}(\varphi l_{\nu}l_{\rho}) + \wt{\nabla}_{\nu}(\varphi l_{\mu}l_{\rho}) - \wt{\nabla}_{\rho}(\varphi l_{\mu}l_{\nu})\right)\, .
\ee
Hence, the Einstein equation is equivalent to $R^{(1)}_{\mu\nu} = 0$.

\subsection{The Kerr-Schild ansatz in Double Field Theory}

A generalization of the Kerr-Schild ansatz to Double Field Theory was introduced in \cite{Lee:2018gxc}. This generalization consists of a perturbative ansatz for the generalized metric in terms of a pair of $O(D,D)$ vectors
\be
\label{gks-ansatz-metric}
H_{MN} = \wt{H}_{MN} + \kappa\varphi\left(K_{M}\bar{K}_{N} + \bar{K}_{M}K_{N}\right)\, .
\ee
As in general relativity, $\kappa$ is an expansion parameter with no physical significance, while $\varphi$ is a scalar function but now depending on the double coordinates. The background generalized metric $\wt{H}_{MN}$ is constrained by the condition
\be
\label{dft-constraint}
\eta_{MN} = \wt{H}_{MP}\eta^{PQ}\wt{H}_{QN}\, .
\ee
The vectors $K$ and $\bar{K}$ satisfy the nullity conditions
\be
\label{gks-nullity}
K_{M}K^{M} = 0\, , \qquad \bar{K}_{M}\bar{K}^{M} = 0\, , 
\ee
and the background chirality conditions
\be
\label{gks-backg-chirality}
\wt{P}_{MN}K^{N} = K_{M}\, , \qquad \bar{\wt{P}}_{MN}\bar{K}^{N} = \bar{K}_{M}\, , \qquad K_{M}\bar{K}^{M} = 0\, ,
\ee
where $\wt{P}$ and $\bar{\wt{P}}$ are the background projectors. Under these conditions, the metric ansatz \eqref{gks-ansatz-metric} automatically satisfies the $O(D,D)$ constraint \eqref{dft-constraint} without requiring any approximations or truncations. Furthermore, the chirality conditions \eqref{gks-backg-chirality} remain valid when the background projectors are replaced by the projectors $P$ and $\bar{P}$ defined in terms of \eqref{gks-ansatz-metric}.

As we aim to incorporate both supersymmetry and $\alpha'$-corrections, it is necessary to perturb the generalized frame. A proposal satisfying the constraints \eqref{dft-frame-constraints}, and consistent with \eqref{gks-ansatz-metric} is
\bea
\label{gks-ansatz-frame-un}
E_{M}{}^{\un{a}} & = & \wt{E}_{M}{}^{\un{a}} - \frac{1}{2}\kappa\varphi \bar{K}_{\ov{b}}K^{\un{a}}\wt{E}_{M}{}^{\ov{b}}\, , \\
\label{gks-ansatz-frame-ov}
E_{M}{}^{\ov{a}} & = & \wt{E}_{M}{}^{\ov{a}} + \frac{1}{2}\kappa\varphi K_{\un{b}}\bar{K}{}^{\ov{a}}\wt{E}_{M}{}^{\un{b}}\, . 
\eea
Here $\wt{E}_{M}{}^{A}$ is the background frame, and the flat vectors are defined as
\be
\label{gks-flat-vectors}
K^{\un{a}} = \wt{E}_{M}{}^{\un{a}}K^{M}\, , \qquad \bar{K}^{\ov{a}} = \wt{E}_{M}{}^{\ov{a}}\bar{K}^{M}\, .
\ee

Before studying the field equations, we need to complete the analysis of the field content of DFT by examining the role of the generalized dilaton in the presence of the Kerr-Schild ansatz. The DFT dilaton $d$ is not subject to any restrictions and does not necessarily have to be linear. The proposal introduced in \cite{Lee:2018gxc} follows
\be
\label{gks-dilaton}
d = \wt{d} + \sum_{n=1}^{\infty}\kappa^{n}d^{(n)}\, , 
\ee
where $\wt{d}$ is the background dilaton and $d^{(n)}$ are perturbations that may depend on the coordinates. For our purposes, we consider only constant contributions to $d^{(n)}$. 

Looking at the equations of motion \eqref{dft-EOM's-frame}, it is essential to examine how the perturbation ansatz affects the generalized fluxes \eqref{dft-generalized-fluxes}. As a preliminary step, it is useful to list the following constraints derived from the nullity condition \eqref{gks-nullity}
\be
\label{gks-diff-null}
K^{\un{a}}\partial_{M}K_{\un{a}} = 0\, , \qquad \bar{K}^{\ov{a}}\partial_{M}\bar{K}_{\ov{a}} = 0\, .
\ee
Additionally, we consider the DFT generalization of the geodesic condition \eqref{gr-geodesic}
\bea
\label{gks-part-geodesic}
K^{\un{a}}\wt{\nabla}_{\un{a}}\bar{K}_{\ov{b}} = 0 & \ \longrightarrow \ & K^{\un{a}}\partial_{\un{a}}\bar{K}_{\ov{b}} = \wt{F}_{\un{a}\ov{bc}}K^{\un{a}}\bar{K}^{\ov{c}}\, , \\ \bar{K}^{\ov{a}}\wt{\nabla}_{\ov{a}}K_{\un{b}} = 0 & \ \longrightarrow \ & \bar{K}^{\ov{a}}\partial_{\ov{a}}K_{\un{b}} = \wt{F}_{\ov{a}\un{bc}}\bar{K}^{\ov{a}}K^{\un{c}}\, , \nn
\eea
where we introduce the background flat derivative $\partial_{A}=\sqrt{2}\wt{E}^{M}{}_{A}\partial_{M}$.

As shown in \cite{Lee:2018gxc}, the equations of motion obtained from the generalized Kerr-Schild ansatz include quadratic terms in the perturbation parameter. However, these terms vanish when the relations \eqref{gks-part-geodesic} are imposed, along with the assumption that the perturbations of the generalized dilaton remain constant. Considering the previous conditions the generalized flux components required to analyze the equations of motion becomes
\bea
\label{gks-flux-one-un}
F_{\un{a}} & = & \wt{F}_{\un{a}} - \frac{1}{2}\kappa\partial_{\ov{b}}\left(\varphi\bar{K}{}^{\ov{b}}K_{\un{a}}\right) - \frac{1}{2}\kappa\varphi\bar{K}{}^{\ov{b}}K_{\un{a}}\wt{F}_{\ov{b}}\, , \\
\label{gks-flux-three-un}
F_{\un{abc}} & = & \wt{F}_{\un{abc}} - \frac{3}{2}\kappa\varphi\bar{K}^{\ov{d}}K_{[\un{a}}\wt{F}_{\un{bc}]\ov{d}}\, , \\
\label{gks-flux-three-ovunun}
F_{\ov{a}\un{bc}} & = & \wt{F}_{\ov{a}\un{bc}} - \kappa\partial_{[\un{b}}\left(\varphi K_{\un{c}]}\bar{K}_{\ov{a}}\right) - \kappa\varphi \bar{K}^{\ov{d}}K_{[\un{b}}\wt{F}_{\un{c}]\ov{ad}} + \frac{\kappa\varphi}{2}K{}^{\un{d}}\bar{K}_{\ov{a}}\wt{F}_{\un{dbc}}\, , \\
\label{gks-flux-three-unovov}
F_{\un{a}\ov{bc}} & = & \wt{F}_{\un{a}\ov{bc}} + \kappa\partial_{[\ov{b}}\left(\varphi\bar{K}_{\ov{c}]}K_{\un{a}}\right) + \kappa\varphi K^{\un{d}}\bar{K}_{[\ov{b}}\wt{F}_{\ov{c}]\un{ad}} - \frac{\kappa\varphi}{2}\bar{K}{}^{\ov{d}}K_{\un{a}}\wt{F}_{\ov{dbc}}\, .
\eea

We can now express the DFT equations of motion in an arbitrary curved background using the generalized Kerr-Schild. The equation of motion of the generalized dilaton is given by
\bea
\label{gks-eom-dilaton}
{\cal R} & = & \wt{{\cal R}} - \kappa\partial_{\un{a}}\partial_{\ov{b}}\left(\varphi K^{\un{a}}\bar{K}{}^{\ov{b}}\right) - \kappa\partial_{\un{a}}\left(\varphi K^{\un{a}}\bar{K}{}^{\ov{b}}\right)\wt{F}_{\ov{b}} - \kappa\varphi K^{\un{a}}\bar{K}{}^{\ov{b}}\partial_{\un{a}}\wt{F}_{\ov{b}}\, \\
& & - \kappa\varphi K^{\un{a}}\bar{K}{}^{\ov{b}}\partial_{\ov{b}}\wt{F}_{\un{a}} - \kappa\partial_{\ov{b}}\left(\varphi K^{\un{a}}\bar{K}{}^{\ov{b}}\right)\wt{F}_{\un{a}} - \kappa\varphi K^{\un{a}}\bar{K}{}^{\ov{b}}\wt{F}_{\un{a}}\wt{F}_{\ov{b}}\, \nn \\
& & - \frac{1}{2}\kappa\varphi K^{\un{a}}\bar{K}^{\ov{b}}\wt{F}_{\un{a}}{}^{\un{cd}}\wt{F}_{\ov{b}\un{cd}} + 2\kappa\partial_{\un{b}}\left(\varphi K_{\un{c}}\bar{K}_{\ov{a}}\right)\wt{F}^{\ov{a}\un{bc}} + 2\kappa\varphi \bar{K}^{\ov{d}}K_{\un{b}}\wt{F}_{\un{c}\ov{ad}}\wt{F}^{\ov{a}\un{bc}}\, \nn ,
\eea
while for the generalized frame, the equation of motion can be written schematically as ${\cal R}_{\ov{a}\un{b}} = \wt{{\cal R}}_{\ov{a}\un{b}} + \kappa{\cal R}^{(1)}_{\ov{a}\un{b}} + \kappa^{2}{\cal R}^{(2)}_{\ov{a}\un{b}}$, with

\bea
\label{gks-eom-frame-1}
{\cal R}^{(1)}_{\ov{a}\un{b}} & = & - \frac{1}{2}\partial_{\ov{a}}\partial_{\ov{c}}\left(\varphi\bar{K}{}^{\ov{c}}K_{\un{b}}\right) + \frac{1}{2}\partial_{\un{c}}\partial_{\un{b}}\left(\varphi K^{\un{c}}\bar{K}_{\ov{a}}\right) - \frac{1}{2}\partial_{\un{c}}\partial^{\un{c}}\left(\varphi K_{\un{b}}\bar{K}_{\ov{a}}\right) - \frac{1}{2}\partial_{\ov{a}}\left(\varphi\bar{K}{}^{\ov{c}}K_{\un{b}}\right)\wt{F}_{\ov{c}}\, \nn \\ 
& & + \frac{1}{2}\partial_{\un{b}}\left(\varphi K_{\un{c}}\bar{K}_{\ov{a}}\right)\wt{F}^{\un{c}} - \frac{1}{2}\partial_{\un{c}}\left(\varphi K_{\un{b}}\bar{K}_{\ov{a}}\right)\wt{F}^{\un{c}} - \frac{1}{2}\partial_{\un{c}}\left(\varphi \bar{K}^{\ov{d}}K^{\un{c}}\right)\wt{F}_{\un{b}\ov{ad}} - \frac{1}{2}\partial_{\un{c}}\left(\varphi K{}^{\un{d}}\bar{K}_{\ov{a}}\right)\wt{F}_{\un{db}}{}^{\un{c}}\, \nn \\
& & - \frac{1}{2}\partial_{\un{b}}\left(\varphi K^{\un{c}}\bar{K}^{\ov{d}}\right)\wt{F}_{\un{c}\ov{da}} + \frac{1}{2}\partial_{\ov{d}}\left(\varphi\bar{K}_{\ov{a}}K_{\un{c}}\right)\wt{F}_{\un{b}}{}^{\un{c}\ov{d}} - \frac{1}{2}\partial_{\ov{a}}\left(\varphi\bar{K}_{\ov{d}}K_{\un{c}}\right)\wt{F}_{\un{b}}{}^{\un{c}\ov{d}} + \frac{1}{2}\partial_{\ov{d}}\left(\varphi\bar{K}{}^{\ov{d}}K^{\un{c}}\right)\wt{F}_{\ov{a}\un{bc}}\, \nn \\
& & - \frac{1}{2}\varphi\bar{K}{}^{\ov{c}}K_{\un{b}}\partial_{\ov{a}}\wt{F}_{\ov{c}} + \frac{1}{2}\varphi K^{\un{c}}\bar{K}{}_{\ov{a}}\partial_{\un{c}}\wt{F}_{\un{b}} + \frac{1}{2}\varphi\bar{K}^{\ov{d}}K_{\un{b}}\partial_{\un{c}}\wt{F}_{\ov{ad}}{}^{\un{c}} - \frac{1}{2}\varphi\bar{K}^{\ov{d}}K^{\un{c}}\partial_{\un{c}}\wt{F}_{\un{b}\ov{ad}} - \frac{1}{2}\varphi K{}^{\un{d}}\bar{K}_{\ov{a}}\partial_{\un{c}}\wt{F}_{\un{db}}{}^{\un{c}}\, \nn \\
& & + \frac{1}{2}\varphi \bar{K}^{\ov{d}}K^{\un{c}}\partial_{\ov{d}}\wt{F}_{\ov{a}\un{bc}} + \frac{1}{2}\varphi \bar{K}^{\ov{d}}K_{\un{b}}\wt{F}_{\un{c}\ov{ad}}\wt{F}^{\un{c}} - \frac{1}{2}\varphi \bar{K}^{\ov{d}}K_{\un{c}}\wt{F}_{\un{b}\ov{ad}}\wt{F}^{\un{c}} - \frac{1}{2}\varphi K{}^{\un{d}}\bar{K}_{\ov{a}}\wt{F}_{\un{dbc}}\wt{F}^{\un{c}}\, \nn \\
& & + \frac{1}{2}\varphi\bar{K}{}^{\ov{d}}K^{\un{c}}\wt{F}_{\ov{a}\un{bc}}\wt{F}_{\ov{d}} - \frac{1}{2}\varphi \bar{K}_{\ov{e}}K_{\un{b}}\wt{F}^{\un{c}\ov{de}}\wt{F}_{\un{c}\ov{da}} + \frac{1}{2}\varphi \bar{K}_{\ov{e}}K^{\un{c}}\wt{F}_{\un{b}}{}^{\ov{de}}\wt{F}_{\un{c}\ov{da}} + \frac{1}{2}\varphi K{}^{\un{e}}\bar{K}^{\ov{d}}\wt{F}_{\un{eb}}{}^{\un{c}}\wt{F}_{\un{c}\ov{da}}\, \nn \\
& & + \frac{1}{2}\varphi K^{\un{e}}\bar{K}_{\ov{d}}\wt{F}_{\ov{a}\un{ce}}\wt{F}_{\un{b}}{}^{\un{c}\ov{d}} - \frac{1}{2}\varphi K^{\un{e}}\bar{K}_{\ov{a}}\wt{F}_{\ov{d}\un{ce}}\wt{F}_{\un{b}}{}^{\un{c}\ov{d}} - \frac{1}{2}\varphi\bar{K}{}^{\ov{e}}K_{\un{c}}\wt{F}_{\ov{eda}}\wt{F}_{\un{b}}{}^{\un{c}\ov{d}}\, ,
\eea
\bea
\label{gks-eom-frame-2}
{\cal R}^{(2)}_{\ov{a}\un{b}} & = & - \frac{1}{4}\partial_{\ov{d}}\left(\varphi\bar{K}_{\ov{a}}K_{\un{c}}\right)\partial_{\un{b}}\left(\varphi K^{\un{c}}\bar{K}^{\ov{d}}\right) + \frac{1}{4}\partial_{\ov{d}}\left(\varphi\bar{K}_{\ov{a}}K_{\un{c}}\right)\partial^{\un{c}}\left(\varphi K_{\un{b}}\bar{K}^{\ov{d}}\right) + \frac{1}{4}\partial_{\ov{a}}\left(\varphi\bar{K}_{\ov{d}}K_{\un{c}}\right)\partial_{\un{b}}\left(\varphi K^{\un{c}}\bar{K}^{\ov{d}}\right)\, \nn \\
& & - \frac{1}{4}\partial_{\ov{a}}\left(\varphi\bar{K}_{\ov{d}}K_{\un{c}}\right)\partial^{\un{c}}\left(\varphi K_{\un{b}}\bar{K}^{\ov{d}}\right) + \frac{1}{4}\partial_{\ov{d}}\left(\varphi\bar{K}{}^{\ov{d}}K^{\un{c}}\right)\partial_{\un{c}}\left(\varphi K_{\un{b}}\bar{K}_{\ov{a}}\right) - \frac{1}{4}\partial_{\ov{d}}\left(\varphi\bar{K}{}^{\ov{d}}K^{\un{c}}\right)\partial_{\un{b}}\left(\varphi K_{\un{c}}\bar{K}_{\ov{a}}\right)\, \nn \\
& & - \frac{1}{4}\varphi K^{\un{d}}\bar{K}{}_{\ov{a}}\partial_{\un{d}}\partial_{\ov{c}}\left(\varphi\bar{K}{}^{\ov{c}}K_{\un{b}}\right) - \frac{1}{4}\varphi \bar{K}^{\ov{d}}K^{\un{c}}\partial_{\ov{d}}\partial_{\un{b}}\left(\varphi K_{\un{c}}\bar{K}_{\ov{a}}\right) + \frac{1}{4}\varphi \bar{K}^{\ov{d}}K^{\un{c}}\partial_{\ov{d}}\partial_{\un{c}}\left(\varphi K_{\un{b}}\bar{K}_{\ov{a}}\right)\, \nn \\
& & - \frac{1}{4}\varphi \bar{K}^{\ov{e}}K^{\un{c}}\partial_{\ov{e}}\left(\varphi \bar{K}^{\ov{d}}K_{\un{b}}\right)\wt{F}_{\un{c}\ov{ad}} - \frac{1}{4}\varphi \bar{K}_{\ov{e}}K_{\un{b}}\partial_{\ov{d}}\left(\varphi\bar{K}_{\ov{a}}K_{\un{c}}\right)\wt{F}^{\un{c}\ov{de}} + \frac{1}{4}\varphi\bar{K}_{\ov{e}}K_{\un{b}}\partial_{\ov{a}}\left(\varphi\bar{K}_{\ov{d}}K_{\un{c}}\right)\wt{F}^{\un{c}\ov{de}}\, \nn \\
& & + \frac{1}{4}\varphi K^{\un{e}}\bar{K}_{\ov{a}}\partial_{\un{b}}\left(\varphi K^{\un{c}}\bar{K}^{\ov{d}}\right)\wt{F}_{\ov{d}\un{ce}} - \frac{1}{4}\varphi K^{\un{e}}\bar{K}_{\ov{a}}\partial^{\un{c}}\left(\varphi K_{\un{b}}\bar{K}^{\ov{d}}\right)\wt{F}_{\ov{d}\un{ce}} - \frac{1}{4}\varphi\bar{K}{}^{\ov{e}}K_{\un{c}}\partial^{\un{c}}\left(\varphi K_{\un{b}}\bar{K}^{\ov{d}}\right)\wt{F}_{\ov{eda}}\, \nn \\
& & - \frac{1}{4}\varphi \bar{K}^{\ov{d}}K_{\un{b}}\partial_{\ov{e}}\left(\varphi\bar{K}{}^{\ov{e}}K^{\un{c}}\right)\wt{F}_{\un{c}\ov{ad}} - \frac{1}{4}\varphi K^{\un{d}}\bar{K}{}_{\ov{a}}\partial_{\un{d}}\left(\varphi\bar{K}{}^{\ov{c}}K_{\un{b}}\right)\wt{F}_{\ov{c}} + \frac{1}{4}\varphi\bar{K}{}^{\ov{e}}K^{\un{c}}\partial_{\un{c}}\left(\varphi K_{\un{b}}\bar{K}_{\ov{a}}\right)\wt{F}_{\ov{e}}\, \nn \\
& & - \frac{1}{4}\varphi^{2}K^{\un{d}}\bar{K}{}_{\ov{a}}\bar{K}{}^{\ov{c}}K_{\un{b}}\partial_{\un{d}}\wt{F}_{\ov{c}} - \frac{1}{4}\varphi^{2} \bar{K}^{\ov{e}}K^{\un{c}}\bar{K}^{\ov{d}}K_{\un{b}}\partial_{\ov{e}}\wt{F}_{\un{c}\ov{ad}} + \frac{1}{4}\varphi^{2}K^{\un{e}}\bar{K}_{\ov{a}}\bar{K}_{\ov{f}}K_{\un{b}}\wt{F}_{\ov{d}\un{ce}}\wt{F}^{\un{c}\ov{df}}\, \nn \\
& & + \frac{1}{4}\varphi^{2}\bar{K}{}^{\ov{e}}K_{\un{c}}\bar{K}_{\ov{f}}K_{\un{b}}\wt{F}_{\ov{eda}}\wt{F}^{\un{c}\ov{df}} - \frac{1}{4}\varphi^{2}\bar{K}{}^{\ov{e}}K^{\un{c}}\bar{K}^{\ov{d}}K_{\un{b}}\wt{F}_{\un{c}\ov{ad}}\wt{F}_{\ov{e}}\, .
\eea

We know that ${\cal R}_{\ov{a}\un{b}} = 0$ holds, and hence, a consistency condition is 
\be
\bar{K}{}^{\ov{a}}K^{\un{b}}{\cal R}_{\ov{a}\un{b}} = \bar{K}{}^{\ov{a}}K^{\un{b}}\wt{{\cal R}}_{\ov{a}\un{b}} + \kappa\bar{K}{}^{\ov{a}}K^{\un{b}}{\cal R}^{(1)}_{\ov{a}\un{b}} + \kappa^{2}\bar{K}{}^{\ov{a}}K^{\un{b}}{\cal R}^{(2)}_{\ov{a}\un{b}} = 0\, ,
\ee
with each of the three terms vanishing. It is straightforward to see that this condition is satisfied by the expansion.

\section{$\alpha'$-corrections and the generalized Kerr-Schild ansatz}

Higher-derivative interactions in the effective field theories describing massless string states are strongly constrained by spacetime symmetries and dualities. Within the framework of DFT, one can systematically construct a series of $\alpha'$-corrections to the action \eqref{dft-action} by extending both the duality group and the double Lorentz group. Identifying the new gauge vectors with the generalized spin connection leads to the following action \cite{Baron:2018lve, Baron:2020xel}
\be
\label{dft-first-order-action}
S = \int d^{2D}X \, e^{-2d} \left( {\cal R} + a{\cal R}^{(-)} + b{\cal R}^{(+)} + {\cal O}(\alpha'{}^{2})\right)\, ,
\ee
where ${\cal R}$ denotes the generalized Ricci scalar \eqref{dft-generalized_ricci}, ${\cal R}^{(\pm)}$ are four-derivative terms constructed from generalized fluxes, and $a$, $b$ are parameters proportional to $\alpha'$ that interpolate between different theories depending on their values\footnote{The full gauge-invariant action is presented in Appendix \ref{appendix}.}. The extension of the duality group requires higher-order corrections to the Lorentz transformation of the generalized frame. At first order, this induces a generalization of the Green-Schwarz mechanism \cite{Marques:2015vua} for anomaly cancellation, given by
\be
\label{ap-generalized-gs}
\delta^{(1)}E_{M}{}^{A} = \left(a\partial_{[\un{M}}\Lambda^{\un{cd}}F_{\ov{N]}\un{cd}} - b\partial_{[\ov{M}}\Lambda^{\ov{cd}}F_{\un{N]}\ov{cd}}\right)E^{N A}\, .
\ee

In what follows, we focus on the four-derivative sector and analyze the implications of applying the gKSA. Our starting point is the perturbation of the generalized Green–Schwarz transformation \eqref{ap-generalized-gs}, from which we derive a set of constraints that propagate into the $\alpha'$-corrected equations of motion of DFT. This extends the analysis of \cite{Lescano:2021ooe} by considering the full bi-parametric theory of \cite{Marques:2015vua}, rather than only the heterotic sector corresponding to $(a=0, b=\alpha')$.

\subsection{Perturbing the generalized Green-Schwarz transformations}

To investigate the interplay between $\alpha'$-corrections and the gKSA, we begin by perturbing the generalized Green–Schwarz transformation. Starting from the first-order correction \eqref{ap-generalized-gs}, we expand both projections of the generalized frame according to \eqref{gks-ansatz-frame-un}-\eqref{gks-ansatz-frame-ov}, and compute the deformation separately in each sector.

As a first step, we rewrite the transformation for both components using only flat indices
\bea
\label{gks-full-gs-un}
\delta^{(1)}E_{M}{}^{\un{a}} & = & \frac{1}{4}\left(a D^{\un{a}}\Lambda^{\un{cd}}F_{\ov{b}\un{cd}} + b D_{\ov{b}}\Lambda_{\ov{cd}}F^{\un{a}\ov{cd}}\right)E_{M}{}^{\ov{b}}\, , \\
\label{gks-full-gs-ov}
\delta^{(1)}E_{M}{}^{\ov{a}} & = & - \frac{1}{4}\left(a D_{\un{b}}\Lambda_{\un{cd}}F^{\ov{a}\un{cd}} + b D^{\ov{a}}\Lambda^{\ov{cd}}F_{\un{b}\ov{cd}}\right)E_{M}{}^{\un{b}}\, ,
\eea
together with their background counterparts
\bea
\delta^{(1)}\wt{E}_{M}{}^{\un{a}} & = & \frac{1}{4}\left(a \partial^{\un{a}}\Lambda^{\un{cd}}\wt{F}_{\ov{b}\un{cd}} + b \partial_{\ov{b}}\Lambda_{\ov{cd}}\wt{F}^{\un{a}\ov{cd}}\right)\wt{E}_{M}{}^{\ov{b}}\, \nn \\
\delta^{(1)}\wt{E}_{M}{}^{\ov{a}} & = & - \frac{1}{4}\left(a \partial_{\un{b}}\Lambda_{\un{cd}}\wt{F}^{\ov{a}\un{cd}} + \frac{b}{4}\partial^{\ov{a}}\Lambda^{\ov{cd}}\wt{F}_{\un{b}\ov{cd}}\right)\wt{E}_{M}{}^{\un{b}}\, . \nn
\eea
All terms in the transformation share the schematic structure
\be
\delta E = D\Lambda \ F \ E \longrightarrow \delta \left(\wt{E} + E^{(\kappa)}\right) = \left(\partial\Lambda + \partial^{(\kappa)}\Lambda\right)\left(\wt{F} + F^{(\kappa)}\right)\left(\wt{E} + E^{(\kappa)}\right)\, , \nn
\ee
where the label ${(\kappa)}$ denotes linear perturbations. The perturbed fluxes are given in \eqref{gks-flux-one-un}–\eqref{gks-flux-three-unovov}, while the flat derivative also receives corrections through the generalized frame. From this schematic form, one expects contributions up to cubic order in $\kappa$, in addition to the background transformations. However, these cubic terms vanish identically in both projections as a consequence of the null conditions. 

After projecting the indices appropriately, we find
\bea
\kappa\delta^{(1)}\left(\varphi K^{\un{a}}\bar{K}_{\ov{b}}\right) & = & \frac{a}{4}\kappa\varphi K^{\un{a}}\bar{K}^{\ov{e}}\partial_{\ov{e}}\Lambda^{\un{cd}}\wt{F}_{\ov{b}\un{cd}} - \frac{b}{4}\kappa\varphi \bar{K}_{\ov{b}}K^{\un{e}}\partial_{\un{e}}\Lambda_{\ov{cd}}\wt{F}^{\un{a}\ov{cd}}\, \\
& & - \frac{a}{2}\left(\partial^{\un{a}}\Lambda^{\un{cd}} - \frac{1}{2}\kappa\varphi K^{\un{a}}\bar{K}^{\ov{e}}\partial_{\ov{e}}\Lambda^{\un{cd}}\right)F^{(\kappa)}_{\ov{b}\un{cd}}\, \nn \\
& & - \frac{b}{2}\left(\partial_{\ov{b}}\Lambda_{\ov{cd}} + \frac{1}{2}\kappa\varphi \bar{K}_{\ov{b}}K^{\un{e}}\partial_{\un{e}}\Lambda_{\ov{cd}}\right)F^{(\kappa)\un{a}\ov{cd}}\, . \nn
\eea
This expression still contains quadratic terms, which must vanish for consistency. The only way to enforce this is by requiring the following constraints on $K$ and $\bar{K}$
\bea
\label{gks-ap-condition1}
\partial_{[\un{b}}\left(\varphi K_{\un{c}]}\bar{K}_{\ov{a}}\right) + \varphi \bar{K}^{\ov{d}}K_{[\un{b}}\wt{F}_{\un{c}]\ov{ad}} - \frac{\varphi}{2}K{}^{\un{d}}\bar{K}_{\ov{a}}\wt{F}_{\un{dbc}} & = & 0\, , \\
\label{gks-ap-condition2}
\partial_{[\ov{b}}\left(\varphi\bar{K}_{\ov{c}]}K_{\un{a}}\right) + \varphi K^{\un{d}}\bar{K}_{[\ov{b}}\wt{F}_{\ov{c}]\un{ad}} - \frac{\varphi}{2}\bar{K}{}^{\ov{d}}K_{\un{a}}\wt{F}_{\ov{dbc}} & = & 0\, . 
\eea
With these conditions imposed, the Lorentz transformation of the perturbation in the generalized Kerr–Schild ansatz receives $\alpha'$-corrections of the form
\be
\label{gks-ap-pert-transf}
\delta^{(1)}\left(\varphi K^{\un{a}}\bar{K}^{\ov{b}}\right) = \frac{\varphi}{4}\left(a K^{\un{a}}\bar{K}^{\ov{e}}\partial_{\ov{e}}\Lambda_{\un{cd}}\wt{F}^{\ov{b}\un{cd}} - b \bar{K}^{\ov{b}}K^{\un{e}}\partial_{\un{e}}\Lambda_{\ov{cd}}\wt{F}^{\un{a}\ov{cd}}\right)\, .
\ee

Conditions \eqref{gks-ap-condition1}–\eqref{gks-ap-condition2} will prove crucial in the linearization of the $\alpha'$-corrected Killing spinor equations once supersymmetric configurations are considered. Before turning to supersymmetry, however, the next subsection is devoted to analyzing how these conditions restrict the equations of motion at leading order in $\alpha'$.

\subsection{Revisiting the leading order equations of motion}

The action of the gKSA on the generalized Green-Schwarz transformations yields two important results:
\begin{enumerate}
    \item The conditions \eqref{gks-ap-condition1}–\eqref{gks-ap-condition2}, which restrict the perturbations of certain components of the generalized fluxes, and
    \item The first-order transformation of the perturbation, given in \eqref{gks-ap-pert-transf}.
\end{enumerate}

The first point is of particular interest, since it simplifies considerably the order $\ap$ of DFT once the ansatz is imposed, as we will see in the next section for the particular case of the Killing spinor equations. In addition, although these conditions were obtained from a first-order transformation, the constraints themselves are of leading order and therefore may also affect the leading-order quantities. For instance, the effect of these conditions on the leading-order equation of motion for the generalized dilaton \eqref{gks-eom-dilaton}, ${\cal R}=0$, becomes
\bea
{\cal{R}} & = & \wt{{\cal R}} - \kappa\partial_{\un{a}}\partial_{\ov{b}}\left(\varphi K^{\un{a}}\bar{K}{}^{\ov{b}}\right) - \kappa\partial_{\un{a}}\left(\varphi K^{\un{a}}\bar{K}{}^{\ov{b}}\right)\wt{F}_{\ov{b}} - \kappa\varphi K^{\un{a}}\bar{K}{}^{\ov{b}}\partial_{\un{a}}\wt{F}_{\ov{b}} - \kappa\varphi K^{\un{a}}\bar{K}{}^{\ov{b}}\partial_{\ov{b}}\wt{F}_{\un{a}}\, \nn \\
& & - \kappa\partial_{\ov{b}}\left(\varphi K^{\un{a}}\bar{K}{}^{\ov{b}}\right)\wt{F}_{\un{a}} - \kappa\varphi K^{\un{a}}\bar{K}{}^{\ov{b}}\wt{F}_{\un{a}}\wt{F}_{\ov{b}} + \frac{1}{2}\kappa\varphi K^{\un{a}}\bar{K}^{\ov{b}}\wt{F}_{\un{a}}{}^{\un{cd}}\wt{F}_{\ov{b}\un{cd}} \ = \ 0\, ,
\eea
which, using the fact that the background is a solution of the theory, reduces to the linearized equation
\be
\partial_{\un{a}}\partial_{\ov{b}}\left(\varphi K^{\un{a}}\bar{K}{}^{\ov{b}}\right) + \partial_{\un{a}}\left(\varphi K^{\un{a}}\bar{K}{}^{\ov{b}}\wt{F}_{\ov{b}}\right) + \partial_{\ov{b}}\left(\varphi K^{\un{a}}\bar{K}{}^{\ov{b}}\wt{F}_{\un{a}}\right) + \varphi K^{\un{a}}\bar{K}{}^{\ov{b}}\wt{F}_{\un{a}}\wt{F}_{\ov{b}} - \frac{1}{2}\varphi K^{\un{a}}\bar{K}^{\ov{b}}\wt{F}_{\un{a}}{}^{\un{cd}}\wt{F}_{\ov{b}\un{cd}} = 0\, .
\ee

For the generalized frame, the perturbation still produces quadratic contributions 
\be
{\cal R}_{\ov{a}\un{b}} = \wt{{\cal R}}_{\ov{a}\un{b}} + \kappa{\cal R}^{(1)}_{\ov{a}\un{b}} + \kappa^{2}{\cal R}^{(2)}_{\ov{a}\un{b}}\, , \nn
\ee
where, as mentioned, the background satisfies the equations of motion, so $\wt{{\cal R}}_{\ov{a}\un{b}} = 0$. The perturbations then take the form
\bea
{\cal R}^{(1)}_{\ov{a}\un{b}} & = & \frac{1}{2}\varphi\bar{K}_{\ov{a}}K^{\un{c}}\partial_{\un{c}}\wt{F}_{\un{b}} - \frac{1}{2}\partial_{\ov{a}}\partial_{\ov{c}}\left(\varphi\bar{K}{}^{\ov{c}}K_{\un{b}}\right) - \frac{1}{2}\partial_{\ov{a}}\left(\varphi\bar{K}{}^{\ov{c}}K_{\un{b}}\right)\wt{F}_{\ov{c}} - \frac{1}{2}\varphi\bar{K}{}^{\ov{c}}K_{\un{b}}\partial_{\ov{a}}\wt{F}_{\ov{c}}\, \nn \\
& & + \frac{1}{2}\varphi K^{\un{c}}\bar{K}^{\ov{d}}\partial_{\ov{d}}\wt{F}_{\ov{a}\un{bc}} + \frac{1}{2}\partial_{\ov{d}}\left(\varphi\bar{K}{}^{\ov{d}}K^{\un{c}}\right)\wt{F}_{\ov{a}\un{bc}} + \frac{1}{2}\varphi\bar{K}{}^{\ov{d}}K^{\un{c}}\wt{F}_{\ov{d}}\wt{F}_{\ov{a}\un{bc}} \ = \ 0\, ,
\eea
and
\be
{\cal R}^{(2)}_{\ov{a}\un{b}} = - \frac{1}{4}\varphi K^{\un{d}}\bar{K}{}_{\ov{a}}\partial_{\un{d}}\partial_{\ov{c}}\left(\varphi\bar{K}{}^{\ov{c}}K_{\un{b}}\right) - \frac{1}{4}\varphi K^{\un{d}}\bar{K}{}_{\ov{a}}\partial_{\un{d}}\left(\varphi\bar{K}{}^{\ov{c}}K_{\un{b}}\right)\wt{F}_{\ov{c}} - \frac{1}{4}\varphi^{2}K^{\un{d}}\bar{K}{}_{\ov{a}}\bar{K}{}^{\ov{c}}K_{\un{b}}\partial_{\un{d}}\wt{F}_{\ov{c}} \ = \ 0\, .
\ee
As is evident, these expressions are significantly simpler than those obtained without imposing the conditions from the generalized Green–Schwarz transformations, previously presented in \eqref{gks-eom-frame-1}–\eqref{gks-eom-frame-2}.

\section{Supersymmetry}

Supersymmetry plays a crucial role in determining the structure of higher-derivative corrections and constraining the possible deformations of Double Field Theory. As discussed in previous sections, the fermionic fields in the supersymmetric formulation transform as spinors of the local Lorentz group ${\rm O}(9,1)_L$, which in ten dimensions must consistently reduce to a single Lorentz group. This consistency condition eliminates one of the two deformation parameters in the biparametric formulation, leaving only the heterotic contributions $(a=0)$, compatible with $\mathcal{N}=1$ supersymmetry.

The first-order $\alpha'$ corrections to ${\cal N}=1$, $D=10$ supersymmetric DFT were derived in~\cite{Lescano:2021guc}, together with their heterotic parameterization. In that construction, the supersymmetry transformation rules were written in terms of the generalized fluxes~\eqref{dft-generalized-fluxes}, providing a natural framework to explore how the generalized Kerr–Schild ansatz modifies the $\alpha'$-corrected Killing spinor equations. This perspective offers a systematic way to study supersymmetric configurations beyond the leading order and to uncover new structures arising from higher-derivative effects.

In what follows, we apply the gKSA to the $\alpha'$-corrected KSEs in DFT. We start by revisiting the leading-order equations, emphasizing their role in identifying supersymmetric backgrounds and describing how they are affected by the Kerr–Schild deformation. We then present our main results: the form of the KSEs at order~$\alpha'$ for a generic DFT background, and their explicit perturbation under the generalized Kerr–Schild ansatz.

\subsection{Killing spinor equations}

The KSEs consists of conditions ensuring that the supersymmetry variations of the fermionic fields vanish. In DFT, these equations take the form
\be
\nabla_{\ov{a}}\epsilon = 0\, , \qquad \gamma^{\un{a}}\nabla_{\un{a}}\epsilon = 0\, ,
\ee
which correspond to the vanishing of the transformations in \eqref{dft-susy-transformations}. While solutions to the KSEs often satisfy the equations of motion, the reverse is not necessarily true. Solutions that fulfill both are known as supersymmetric solutions.

Similar to how we did for the equations of motion of bosonic field content, we can explore the effect of the gKSA on these equations. The KSE arising from the transformation of the gravitino and dilatino are given by
\be
\label{gks-gravitino-kse}
\nabla_{\ov{a}}\epsilon = D_{\ov{a}}\epsilon + \frac{1}{4}F_{\ov{a}\un{bc}}\gamma^{\un{bc}}\epsilon = 0\, , 
\ee
\be
\label{gks-dilatino-kse}
\gamma^{\un{a}}\nabla_{\un{a}}\epsilon = \gamma^{\un{a}}D_{\un{a}}\epsilon + \frac{1}{12}F_{\un{abc}}\gamma^{\un{abc}}\epsilon + \frac{1}{2}F_{\un{a}}\gamma^{\un{a}}\epsilon = 0\, , 
\ee
where $D_{A}=\sqrt{2}E^{M}{}_{A}\partial_{M}$, with $E^{M}{}_{A}$ being the perturbed frame field, which includes the background frame and thus the flat derivative $\partial_{A}$. 

If $\epsilon$ is a solution to the background Killing Spinor Equations
\be
\wt{\nabla}_{\ov{a}}\epsilon = \partial_{\ov{a}}\epsilon + \frac{1}{4}\wt{F}_{\ov{a}\un{bc}}\gamma^{\un{bc}}\epsilon = 0\, , 
\ee
\be
\gamma^{\un{a}}\wt{\nabla}_{\un{a}}\epsilon = \gamma^{\un{a}}\partial_{\un{a}}\epsilon + \frac{1}{12}\wt{F}_{\un{abc}}\gamma^{\un{abc}}\epsilon + \frac{1}{2}\wt{F}_{\un{a}}\gamma^{\un{a}}\epsilon = 0\, ,
\ee
we can track how the generalized Kerr-Schild perturbation affects the equations. This leads to the general expressions
\be
\label{dft-ksa-kse-gravitino}
\varphi K^{\un{b}}\bar{K}{}_{\ov{a}}\wt{\nabla}_{\un{b}}\epsilon - \frac{1}{2}\wt{\nabla}_{\un{b}}\left(\varphi K_{\un{c}}\bar{K}_{\ov{a}}\right)\gamma^{\un{bc}}\epsilon = 0\, , 
\ee
and 
\be
\label{dft-ksa-kse-dilatino}
\wt{\nabla}_{\ov{b}}\left(\varphi\bar{K}{}^{\ov{b}}K_{\un{a}}\right)\gamma^{\un{a}}\epsilon = 0\, ,
\ee
which coincides with the KSEs obtained in \cite{Lee:2018gxc}. 

However, as we have shown, the perturbation of the generalized Green–Schwarz transformation also imposes constraints on the perturbations of the leading-order generalized fluxes. Consequently, even though at leading order the equations \eqref{dft-ksa-kse-gravitino}–\eqref{dft-ksa-kse-dilatino} accurately describe supersymmetric backgrounds, once $\alpha'$-corrections are included, equation \eqref{dft-ksa-kse-gravitino} yields the additional constraint
\be
\label{dft-ksa-kse-constraint}
K^{\un{b}}\partial_{\un{b}}\epsilon = 0\, .
\ee

\subsection{$\ap$-corrections to the supersymmetry transformations}

The first order $\alpha'$-corrections to the supersymmetry transformations \eqref{dft-susy-transformations} are given by
\be
\label{ap-bosonic-transf}
\delta^{(1)}E_{M}{}^{\un{a}} = \frac{\alpha'}{16}\ov{\epsilon}\gamma^{\un{c}}\Psi_{\ov{b}}F^{*}_{\un{c}\ov{de}}F^{*\un{a}\ov{de}}E_{M}{}^{\ov{b}}\, , \qquad \delta^{(1)}E_{M}{}^{\ov{a}} = - \frac{\alpha'}{16}\ov{\epsilon}\gamma_{\un{c}}\Psi^{\ov{a}}F^{*}_{\un{b}\ov{de}}F^{*\un{c}\ov{de}}E_{M}{}^{\un{b}}\, , 
\ee
\be
\label{ap-fermionic-transf}
\delta^{(1)}\Psi_{\ov{a}} = \frac{1}{4}F^{(3)}_{\ov{a}\un{bc}}\gamma^{\un{bc}}\epsilon\, , \qquad \delta^{(1)}\varrho = - \frac{\alpha'}{8}F^{*}_{\un{a}\ov{cd}}F^{*\un{b}\ov{cd}}\gamma^{\un{a}}D_{\un{b}}\epsilon - \frac{1}{12}F^{(3)}_{\un{abc}}\gamma^{\un{abc}}\epsilon - \frac{1}{2}F^{(3)}_{\un{a}}\gamma^{\un{a}}\epsilon\, ,
\ee
where $F^{(3)}$ denotes the three-derivative corrections to the generalized fluxes. Explicitly
\bea
F^{(3)}_{\ov{a}\un{bc}} & = & - \frac{\alpha'}{4}\left(D_{\ov{a}}F^{*\ov{cd}}_{[\un{b}} +{F}^{*\un{e}\ov{cd}}{F}_{\ov{a}\un{e}[\un{b}}\right)F^{*}_{\un{c}]\ov{cd}}\, , \\
F^{(3)}_{\un{abc}} & = & - \frac{3\alpha'}{4}\left(D_{[\un{a}}F^{*\ov{cd}}_{\un{b}} - \frac{1}{2}F_{\un{d}[\un{ab}}F^{*\un{d}\ov{cd}} - \frac{2}{3}F^{*\ov{c}}{}_{\ov{e}[\un{a}}F^{*}_{\un{b}}{}^{\ov{ed}}\right)F^{*}_{\un{c}]\ov{cd}}\, , \\
F^{(3)}_{\un{a}} & = & \frac{\alpha'}{8}\left[F^{*}_{\un{a}\ov{cd}}F^{*\un{b}\ov{cd}}F_{\un{b}} + D_{\un{b}}\left(F^{*}_{\un{a}\ov{cd}}F^{*\un{b}\ov{cd}}\right) \right]\, ,
\eea
with $F^{*}_{\un{a}\ov{cd}}=F_{\un{a}\ov{cd}} - \frac{1}{2}\ov{\Psi}_{\ov{c}}\gamma_{\un{a}}\Psi_{\ov{d}}$. Since we are working at leading order in the number of fermions, this simplifies to $F^{*}_{\un{a}\ov{cd}}=F_{\un{a}\ov{cd}}$.

To focus on the KSEs, it is convenient to express the supersymmetry transformations of the fermionic fields \eqref{ap-fermionic-transf} in terms of the leading-order generalized fluxes. These take the form
\bea
\label{ap-grav-susy-transf}
\delta\Psi_{\ov{a}} & = & \nabla_{\ov{a}}\epsilon - \frac{\alpha'}{16}D_{\ov{a}}F_{\un{b}}{}^{\ov{cd}}F_{\un{c}\ov{cd}}\gamma^{\un{bc}}\epsilon - \frac{\alpha'}{16}F_{\ov{a}\un{e}\un{b}}F_{\un{c}\ov{cd}}F^{\un{e}\ov{cd}}\gamma^{\un{bc}}\epsilon\, , \\
\label{ap-dil-susy-transf}
\delta\varrho & = & - \gamma^{\un{a}}\nabla_{\un{a}}\epsilon - \frac{\alpha'}{8}F_{\un{a}\ov{cd}}F^{\un{b}\ov{cd}}\gamma^{\un{a}}D_{\un{b}}\epsilon + \frac{\alpha'}{16}D_{\un{a}}F_{\un{b}}{}^{\ov{cd}}F_{\un{c}\ov{cd}}\gamma^{\un{abc}}\epsilon - \frac{\alpha'}{32}F_{\un{d}\un{ab}}F^{\un{d}\ov{cd}}F_{\un{c}\ov{cd}}\gamma^{\un{abc}}\epsilon\, \nn \\
& & - \frac{\alpha'}{24}F^{\ov{c}}{}_{\ov{e}\un{a}}F_{\un{b}}{}^{\ov{ed}}F_{\un{c}\ov{cd}}\gamma^{\un{abc}}\epsilon - \frac{\alpha'}{16}D_{\un{b}}\left(F_{\un{a}\ov{cd}}F^{\un{b}\ov{cd}}\right)\gamma^{\un{a}}\epsilon - \frac{\alpha'}{16}F_{\un{a}\ov{cd}}F^{\un{b}\ov{cd}}F_{\un{b}}\gamma^{\un{a}}\epsilon\, ,
\eea
where the leading order transformations are included. The appearance of higher-derivative terms in the supersymmetry transformations reflects the complexity introduced by the $\alpha'$ expansion, which affects both the bosonic and fermionic sectors of the theory. In particular, these corrections require the inclusion of a Chern–Simons term in the effective action, as demanded by anomaly cancellation in string theory. Hence, the $\ap$-corrected Killing spinor equations becomes
\be
\label{ap-grav-kse}
\nabla_{\ov{a}}\epsilon - \frac{\ap}{16}\left(D_{\ov{a}}F_{\un{b}}{}^{\ov{cd}}F_{\un{c}\ov{cd}} + F_{\ov{a}\un{e}\un{b}}F_{\un{c}\ov{cd}}F^{\un{e}\ov{cd}}\right)\gamma^{\un{bc}}\epsilon = 0\, ,
\ee
\begin{align}
\label{ap-dil-kse}
- \gamma^{\un{a}}\nabla_{\un{a}}\epsilon - \frac{\ap}{8} & \Big[F_{\un{a}\ov{cd}}F^{\un{b}\ov{cd}}\gamma^{\un{a}}D_{\un{b}}\epsilon + \frac{1}{2}\left(D_{\un{b}}\left(F_{\un{a}\ov{cd}}F^{\un{b}\ov{cd}}\right) + F_{\un{a}\ov{cd}}F^{\un{b}\ov{cd}}F_{\un{b}}\right)\gamma^{\un{a}}\epsilon\, \nn \\ 
& - \left(\frac{1}{2}D_{\un{a}}F_{\un{b}}{}^{\ov{cd}} - \frac{1}{4}F_{\un{d}\un{ab}}F^{\un{d}\ov{cd}} - \frac{1}{3}F^{\ov{c}}{}_{\ov{e}\un{a}}F_{\un{b}}{}^{\ov{ed}}\right)F_{\un{c}\ov{cd}}\gamma^{\un{abc}}\epsilon\Big] \  = \ 0\, . 
\end{align}

In what follows, we study these equations within the framework of the generalized Kerr–Schild ansatz, in order to derive suitable conditions that allow for first order supersymmetric solutions.

\subsection{Perturbation of the ${\cal O}(\alpha')$ Killing spinor equations}

The action of the gKSA at order ${\cal O}(\alpha')$ imposes the conditions~\eqref{gks-ap-condition1}–\eqref{gks-ap-condition2}, which are related to the perturbations of specific components of the generalized fluxes. In addition, we have seen that the constraint~\eqref{dft-ksa-kse-constraint} arises from the leading-order KSEs when the gKSA is considered at order~$\alpha'$. These conditions allow for the linearization of the $\alpha'$-corrected KSEs~\eqref{ap-grav-kse}–\eqref{ap-dil-kse}. 

Starting with the KSE associated with the gravitino variation~\eqref{ap-grav-kse}, the gKSA yields the following $\alpha'$-corrected constraint
\be
\label{ap-kse-result1}
\varphi K^{\un{e}}\bar{K}{}_{\ov{a}}\partial_{\un{e}}\wt{F}_{\un{b}}{}^{\ov{cd}}\wt{F}_{\un{c}\ov{cd}}\gamma^{\un{bc}}\epsilon \ = \ 0\, .
\ee

Turning to the dilatino equation~\eqref{ap-dil-kse}, after suitable manipulations and use of the background KSE, the first-order contributions can be written as
\be 
\wt{\nabla}_{\ov{e}}\left(\varphi K_{\un{b}}\bar{K}^{\ov{e}}\wt{F}_{\un{a}\ov{cd}}\wt{F}^{\un{b}\ov{cd}}\right)\gamma^{\un{a}}\epsilon - \varphi K_{\un{a}}\bar{K}^{\ov{e}}\left(\partial_{\ov{e}}\wt{F}_{\un{b}}{}^{\ov{cd}}\wt{F}_{\un{c}\ov{cd}} + \wt{F}_{\ov{e}\un{db}}\wt{F}_{\un{c}\ov{cd}}\wt{F}^{\un{d}\ov{cd}}\right)\gamma^{\un{bca}}\epsilon \ = \ 0\, ,
\ee
where the combination in parentheses in the second term vanishes due to the background $\alpha'$-corrected gravitino KSE. Finally, we obtain
\be 
\label{ap-kse-result2}
- \frac{\ap}{32}\varphi\wt{\nabla}_{\ov{b}}\left(\bar{K}^{\ov{b}}K_{\un{e}}\wt{F}_{\un{a}\ov{cd}}\wt{F}^{\un{e}\ov{cd}}\right)\gamma^{\un{a}}\epsilon \ = \ 0\, ,
\ee
which represents the $\alpha'$-correction to the dilatino Killing spinor equation. It is straightforward to observe that this equation can be absorbed into the leading-order equation~\eqref{gks-dilatino-kse} by redefining the null vector as $K'_{a}=K_{\un{a}}-\frac{\ap}{32}K_{\un{e}}\wt{F}_{\un{a}\ov{cd}}\wt{F}^{\un{e}\ov{cd}}$, a strategy that will be further explored in the next section, once we establish the connection with supergravity.

\section{Supergravity parameterization}

In order to connect the DFT formulation with the low-energy limit of String Theory, it is necessary to express all generalized quantities in terms of conventional supergravity objects: the metric $g_{\mu\nu}$, the Kalb–Ramond two-form $b_{\mu\nu}$, the dilaton $\phi$, and their curvatures. This process requires solving the strong constraint by imposing $\partial^{\mu}=0$, which effectively eliminates the dependence on the dual coordinates and projects the theory onto a $D$-dimensional physical spacetime. In this section we present the explicit parameterization of the fundamental fields and the null geodesic vectors of the gKSA, and illustrate the construction in the context of ten-dimensional ${\cal N}=1$ supergravity of the $\alpha'$-corrections of the KSEs.

\subsection{Fundamental fields and the ansatz}

To extend the DFT results to supergravity, we first parameterize the generalized fields in terms of supergravity fields and solve the strong constraint. The parameterization of the fundamental fields is well known: the generalized metric is given by \eqref{dft-generalized-metric}, and the generalized frame by \eqref{dft-generalized_frame}.

To complete the parameterization of the gKSA, we next express the generalized null vectors $K$ and $\bar{K}$ in terms of $D$-dimensional vectors $l^{\mu}$ and $\bar{l}^{\mu}$ as
\begin{equation}
\label{gks-null-vector-param}
K^{M} = \frac{1}{\sqrt{2}} \begin{pmatrix}
l^{\mu} \\
(-\wt{g}_{\mu\nu} + \wt{b}_{\mu\nu}) l^{\nu}
\end{pmatrix}, \qquad \bar{K}^{M} = -\frac{1}{\sqrt{2}} \begin{pmatrix}
\bar{l}^{\mu} \\
(\wt{g}_{\mu\nu} + \wt{b}_{\mu\nu}) \bar{l}^{\nu}
\end{pmatrix}\, .
\end{equation}
Substituting these into \eqref{gks-nullity} shows that $l^{\mu}$ and $\bar{l}^{\mu}$ are also null with respect to the background metric
\begin{equation}
l^{\mu}\wt{g}_{\mu\nu}l^{\nu} = l^{\mu}l_{\mu} = 0\, , \qquad \bar{l}^{\mu}\wt{g}_{\mu\nu}\bar{l}^{\nu} = \bar{l}^{\mu}\bar{l}_{\mu} = 0\, .
\end{equation}
but, unlike in the conventional Kerr–Schild ansatz, they are not necessarily orthogonal to each other
\be
l^{\mu}\bar{l}_{\mu} \neq 0\, , \qquad l^{\mu}\bar{l}_{\mu} = l \cdot \bar{l} = l^{\mu}\wt{g}_{\mu\nu}\bar{l}^{\nu}\, .
\ee
Here and in what follows, indices are raised and lowered using the background metric $\wt{g}$.

Parameterizing the generalized metric \eqref{gks-ansatz-metric} in terms of the background fields and the vectors \eqref{gks-null-vector-param} yields the corresponding spacetime metric and Kalb–Ramond field
\bea
\label{gks-sugra-metric}
g_{\mu\nu} & = & \wt{g}_{\mu\nu} + \frac{\kappa \varphi}{1 - \frac{1}{2} \kappa \varphi (l \cdot \bar{l})} l_{(\mu} \bar{l}_{\nu)}\, , \\
\label{gks-sugra-b}
b_{\mu\nu} & = & \wt{b}_{\mu\nu} + \frac{\kappa \varphi}{1 - \frac{1}{2} \kappa \varphi (l \cdot \bar{l})} l_{[\mu} \bar{l}_{\nu]}\, ,
\eea
with inverse metric
\be
\label{gks-sugra-metric-inverse}
g^{\mu\nu} = \wt{g}{}^{\mu\nu} - \kappa \varphi l^{(\mu} \bar{l}^{\nu)}\, .
\ee

A key distinction from the conventional Kerr–Schild ansatz is that this generalized construction induces perturbations in both the metric and the Kalb–Ramond field due to the presence of two independent null vectors $l^{\mu}$ and $\bar{l}^{\mu}$, rather than a single one. Moreover, these vectors satisfy
\be
l_\mu g^{\mu\nu} \bar{l}_\nu \neq l_\mu \wt{g}^{\mu\nu} \bar{l}_\nu\, .
\ee
so the generalized and conventional formulations differ in this respect. The two coincide when $l^{\mu} = \bar{l}^{\mu}$, for which $l \cdot \bar{l} \to l^2 = 0$, and the previous expressions reduce to
\be
g_{\mu\nu} = \wt{g}_{\mu\nu} + \kappa\varphi l_{\mu}l_{\nu}\, , \qquad g^{\mu\nu} = \wt{g}^{\mu\nu} - \kappa\varphi l^{\mu}l^{\nu}\, ,
\ee
while the Kalb–Ramond field remains unperturbed. The determinant of the metric takes the form
\be
\det g = (\det \wt{g}) \left( 1 - \frac{1}{2} \kappa \varphi (l \cdot \bar{l}) \right)^{-2}\, ,
\ee
which correctly reproduces the standard Kerr–Schild result in the limit $l = \bar{l}$. It is worth noting that, unlike in the conventional case, the metric and the Kalb–Ramond field in this generalized framework are not linear in $\kappa$.

The parameterization of both components of the generalized frame \eqref{gks-ansatz-frame-un}–\eqref{gks-ansatz-frame-ov} leads to
\be
\label{gks-vielbein}
e_{\mu}{}^{\un{a}} = \wt{e}_{\mu}{}^{\un{a}} + \frac{\kappa\varphi}{2- \kappa\varphi\left(l\cdot\bar{l}\right)}l_{\mu}\bar{l}^{\nu}\wt{e}_{\nu}{}^{\un{a}}\, , \qquad e_{\mu}{}^{\ov{a}} = \wt{e}_{\mu}{}^{\ov{a}} + \frac{\kappa\varphi}{2- \kappa\varphi\left(l\cdot\bar{l}\right)}\bar{l}_{\mu}l^{\nu}\wt{e}_{\nu}{}^{\ov{a}}\, ,
\ee
with inverses
\be
\label{gks-vielbein-inverse}
e^{\mu}{}^{\un{a}} = \wt{e}^{\mu}{}^{\un{a}} - \frac{1}{2}\kappa\varphi \bar{l}^{\mu}l^{\nu}\wt{e}_{\nu}{}^{\un{a}}\, , \qquad e^{\mu}{}^{\ov{a}} = \wt{e}^{\mu}{}^{\ov{a}} - \frac{1}{2}\kappa\varphi l^{\mu}\bar{l}{}^{\nu}\wt{e}_{\nu}{}^{\ov{a}}\, .
\ee
As expected, these vielbeins satisfy
\be
\label{gks-nd_vielbein}
\wt{e}_{\mu}{}^{\un{a}}g_{\un{ab}}\wt{e}_{\nu}{}^{\un{b}} = \wt{e}_{\mu}{}^{\ov{a}}g_{\ov{ab}}\wt{e}_{\nu}{}^{\ov{b}} = \wt{g}_{\mu\nu}\, , \qquad e_{\mu}{}^{\un{a}}g_{\un{ab}}e_{\nu}{}^{\un{b}} = e_{\mu}{}^{\ov{a}}g_{\ov{ab}}e_{\nu}{}^{\ov{b}} = g_{\mu\nu}\, , 
\ee
which, as discussed earlier, must be gauge-fixed to identify the physical supergravity vielbein and reduce the double Lorentz group to the standard Lorentz group.

At the supergravity level, the dilaton takes the form
\be
\phi = \wt{\phi} + \delta\phi\, , \qquad \delta\phi = \text{const.}\, ,
\ee
where the constant shift compensates for the deformation of the metric determinant, ensuring that the integration measure remains invariant under the gKSA. This completes the parameterization of the fundamental fields.

It is also useful to write the background generalized fluxes, which play a central role in both the field equations and the Killing spinor equations. Their parameterization is
\bea
\wt{F}_{\un{a}} & = & \partial_{\mu}\wt{e}^{\mu}{}_{a} + \wt{e}^{\nu}{}_{b}\partial_{a}\wt{e}_{\nu}{}^{b} - 2\partial_{a}\wt{\phi}\, , \\
\wt{F}_{\ov{a}} & = & \partial_{\mu}\wt{e}^{\mu}{}_{a} + \wt{e}^{\nu}{}_{b}\partial_{a}\wt{e}_{\nu}{}^{b} - 2\partial_{a}\wt{\phi}\, , \\
\wt{F}_{\un{abc}} & = & -3\left(\wt{w}_{[abc]} + \frac{1}{6}\wt{H}_{abc}\right)\, , \\
\wt{F}_{\ov{abc}} & = & 3\left(\wt{w}_{[abc]} - \frac{1}{6}\wt{H}_{abc}\right)\, , \\
\wt{F}_{\ov{a}\un{bc}} & = & -\left(\wt{w}_{abc} + \frac{1}{2}\wt{H}_{abc}\right) = - \wt{\Omega}{}^{(+)}_{abc}\, , \\
\wt{F}_{\un{a}\ov{bc}} & = & \wt{w}_{abc} - \frac{1}{2}\wt{H}_{abc} = \wt{\Omega}{}^{(-)}_{abc}\, .
\eea
Here $\wt{w}_{abc}$ denotes the background spin connection
\be
\wt{w}_{abc} = -\wt{e}^{\mu}{}_{[a}\wt{e}^{\nu}{}_{b]}\partial_{\mu}\wt{e}_{\nu c} + \wt{e}^{\mu}{}_{[a}\wt{e}^{\nu}{}_{c]}\partial_{\mu}\wt{e}_{\nu b} + \wt{e}^{\mu}{}_{[b}\wt{e}^{\nu}{}_{c]}\partial_{\mu}\wt{e}_{\nu a}\, ,
\ee
and $\wt{H}_{abc}$ is the field strength of the Kalb–Ramond field
\be
\wt{H}_{abc} = \wt{e}^{\mu}{}_{a}\wt{e}^{\nu}{}_{b}\wt{e}^{\rho}{}_{c}\wt{H}_{\mu\nu\rho}\, , \qquad \wt{H}_{\mu\nu\rho} = 3\partial_{[\mu}\wt{b}_{\nu\rho]}\, . 
\ee
We also define the torsionful spin connections
\be
\wt{\Omega}^{(\pm)}_{abc} = \wt{w}_{abc} \pm \frac{1}{2}\wt{H}_{abc}\, .
\ee

Finally, the flat generalized null vectors introduced in \eqref{gks-flat-vectors} read
\be
K^{\un{a}} = - \wt{e}_{\mu}{}^{\un{a}}l^{\mu} = - l^{\un{a}}\, , \qquad \bar{K}^{\ov{a}} = -\wt{e}_{\mu}{}^{\ov{a}}\bar{l}^{\mu} = -\bar{l}^{\ov{a}}\, ,
\ee
which will be used to express the relevant quantities of interest, such as the constraints and the KSEs, in a compact and manifestly flat-index form.

\subsection{$D=10$, ${\cal N}=1$ supergravity at ${\cal O}(\ap)$: the Killing spinor equations}

At this point, we finally have all the ingredients to write down the necessary conditions to search for supersymmetric backgrounds satisfying the generalized Kerr–Schild ansatz.

Let us start with the parameterization of the transformation \eqref{gks-ap-pert-transf} for the choice of parameters $(a=0,b=\ap)$, which becomes
\be
\label{param-gs}
\delta_{\Lambda}^{(1)}\left(\varphi l_{a}\bar{l}_{b}\right) = \frac{\ap\varphi}{4} \bar{l}_{b}l^{e}\partial_{e}\Lambda^{cd}\wt{\Omega}^{(-)}_{acd}\, ,
\ee
compensating the extra contributions coming from the perturbation of the vielbein so as to preserve the Green–Schwarz transformation for both the background and the perturbed vielbein. As we have shown, in order to reach this transformation, it is necessary to impose the conditions \eqref{gks-ap-condition1}–\eqref{gks-ap-condition2}, whose parameterization reads
\bea
\label{param-cond1}
& & \wt{\nabla}_{[b}\left(\varphi l_{c]}\bar{l}_{a}\right) - \frac{\varphi}{2}\bar{l}^{d}l_{[b}\wt{H}_{c]ad} - \frac{\varphi}{2}l^{d}\bar{l}_{a}\wt{\Omega}^{(+)}_{dbc} = 0 \, , \\
\label{param-cond2}
& & \nabla_{[b}\left(\varphi\bar{l}_{c]}l_{a}\right) + \frac{1}{2}\varphi l^{d}\bar{l}_{[b}\wt{H}_{c]ad} - \frac{\varphi}{2}\bar{l}^{d}l_{a}\wt{\Omega}^{(-)}_{dbc} = 0\, . 
\eea
These conditions affect the equations of motion, both in the leading and $\ap$-corrected order, and more related to our perspective, they also influence the Killing spinor equations.

From the supersymmetric transformation of the gravitino we obtain the following two constraints
\be
\label{param-constraints}
l^{b}\partial_{b}\epsilon = 0\, , \qquad l^{d}\bar{l}{}_{a}\partial_{d}\wt{\Omega}^{(-)}_{b}{}^{ef}\wt{\Omega}^{(-)}_{cef}\gamma^{bc}\epsilon \ = \ 0\, .
\ee
All these ingredients allow us to write the KSE for the dilatino as
\be 
\label{param-dil-kse}
\wt{\nabla}_{b}\left(\varphi\bar{l}{}^{b}l_{a} + \frac{\ap}{32}\varphi \bar{l}^{b}l_{e}\wt{\Omega}^{(-)}_{acd}\wt{\Omega}^{(-)ecd}\right)\gamma^{a}\epsilon \ = \ 0\, ,
\ee
which can be rewritten in the same form as the leading-order KSE. To achieve this, one must first include the field redefinitions that relate the duality-covariant fields in DFT to those of the supergravity effective action, and then perform a redefinition of the null vector, leading to
\be
\label{param-redefinition}
l'_{\mu} = l_{\mu} + \frac{5\ap}{32}\wt{\Omega}^{(-)}_{\mu cd}\wt{\Omega}^{(-)\nu cd}l_{\nu}\, .
\ee

This redefinition has important implications: it allows supersymmetric solutions at order~$\ap$ to be interpreted as leading-order solutions with the corrected null vector~$l'$ carrying the higher-derivative contributions, provided that the conditions \eqref{param-gs}–\eqref{param-constraints} are satisfied. Moreover, it offers a first glimpse into the structure of higher-derivative corrections, since part of the $\ap$ field redefinitions are already encoded in~$l'$. This shows that the perturbation of leading-order solutions naturally could captures contributions beyond leading order.

\section{Conclusions}

In this work, we have extended the analysis of the generalized Kerr–Schild ansatz (gKSA) in Double Field Theory to explore higher-derivative corrections in the supersymmetric sector. Building upon the $\alpha'$-corrected ten-dimensional $\mathcal{N}=1$ supersymmetric formulation of DFT~\cite{Lescano:2021guc}, we derived the $\mathcal{O}(\alpha')$ Killing spinor equations (KSEs) and studied how they are modified under the action of the gKSA.

A key outcome of our analysis is that the constraints \eqref{gks-ap-condition1}–\eqref{gks-ap-condition2}, arising from the perturbation of the generalized Green–Schwarz transformations, play a fundamental role in the supersymmetric theory. These constraints ensure the linearization of the $\alpha'$-corrected KSEs, providing a systematic framework to analyze higher-derivative supersymmetric configurations. We have also shown that the $\alpha'$-corrected KSEs can be expressed compactly in terms of the background generalized fluxes and the null vectors defining the ansatz. When rewritten in the supergravity parameterization, these equations yield an effective first-order description of supersymmetric backgrounds, bridging the gap between the duality-invariant formalism and conventional $\mathcal{N}=1$ supergravity.

The ${\cal O}(\alpha')$ equations of motion of heterotic DFT were previously obtained in~\cite{Lescano:2021ooe}. However, those equations are highly intricate, making the search for explicit solutions extremely challenging. We expect that the results of this work will significantly simplify their structure, rendering the equations more tractable and facilitating the construction of explicit $\alpha'$-corrected supersymmetric solutions. A natural continuation of this work would be to apply the gKSA to the full bi-parametric theory of~\cite{Marques:2015vua}, obtaining perturbed equations of motion that depend on two deformation parameters specifying the effective theory under consideration. While this generalized framework does not admit a supersymmetric extension—since neither the bosonic theory nor the HSZ formulation are supersymmetric—it may still reveal interesting structural insights.

An intriguing direction concerns the relation between the generalized Kerr–Schild ansatz and the classical double copy \cite{Lee:2018gxc, Cho:2019ype, Kim:2019jwm, Lescano:2020nve, Berman:2020xvs, Lescano:2021ooe, Angus:2021zhy, Lescano:2022nhp, Berman:2022bgy}. The gKSA has already proven to be a powerful tool for uncovering double-copy structures at leading order, and it would be interesting to explore whether similar connections persist in supersymmetric and $\alpha'$-corrected settings. In particular, one could investigate whether the higher-derivative corrections derived here admit a single- or zeroth-copy interpretation, or whether they correspond to the equations of motion of higher-derivative gauge theories related to DFT via double-copy mappings~\cite{Diaz-Jaramillo:2021wtl, Lescano:2023pai, Lescano:2024gma, Lescano:2024lwn}.

Finally, it is worth mentioning that various deformations of DFT fields can lead to different (generalized) supergravity backgrounds. For instance, deformations of the $\beta$ field in $\beta$-supergravity yield backgrounds satisfying the homogeneous Classical Yang–Baxter Equation (CYBE)~\cite{Bakhmatov:2018bvp}. Although the gKSA and $\beta$-deformations arise from different mechanisms, the insights gained from our construction may provide clues for exploring potential connections between integrability and higher-derivative deformations in DFT.

\acknowledgments

The author would like to thank Eric Lescano for valuable comments on the first version of the draft, and in particular for pointing out the importance of considering redefinitions of the vielbein in addition to those of the vector $l'$. Institutional support from the Universidad de Buenos Aires (UBA) and financial support from UADE are also gratefully acknowledged.

\appendix

\section{Gauge Invariant Action}
\label{appendix}

The gauge-invariant DFT action to first order in $\alpha'$ takes the form
\be
S = \int d^{2D}X \, e^{-2d} \left( {\cal R} + a{\cal R}^{(-)} + b{\cal R}^{(+)} + {\cal O}(\alpha'{}^{2})\right)\, , 
\ee
where ${\cal R}$ is the generalized Ricci scalar defined in \eqref{dft-generalized_ricci}. This action is invariant under the standard leading-order gauge symmetries of the theory, supplemented by the generalized Green–Schwarz transformation \eqref{ap-generalized-gs}.  

The first-order contributions are given by
\bea
\label{ap-Rminus}
{\cal R}^{(-)} & = & - \frac{1}{4}D_{\ov{a}}D^{\ov{b}}F^{\ov{a}\un{cd}}F_{\ov{b}\un{cd}} - \frac{1}{4}D_{\ov{a}}D^{\ov{b}}F_{\ov{b}\un{cd}}F^{\ov{a}\un{cd}} + \frac{1}{8}D_{\un{a}}F_{\ov{b}\un{cd}}D^{\un{a}}F^{\ov{b}\un{cd}} - \frac{1}{4}D_{\ov{a}}F^{\ov{a}\un{cd}}D^{\ov{b}}F_{\ov{b}\un{cd}}\, \\
& & - \frac{1}{8}D_{\ov{a}}F_{\ov{b}\un{cd}}D^{\ov{a}}F^{\ov{b}\un{cd}} - \frac{1}{2}D_{\ov{a}}F^{\ov{b}}F^{\ov{a}\un{cd}}F_{\ov{b}\un{cd}} - \frac{1}{4}F_{\ov{a}}F^{\ov{b}}F^{\ov{a}\un{cd}}F_{\ov{b}\un{cd}} - \frac{1}{2}D_{\ov{a}}F^{\ov{a}\un{cd}}F_{\ov{b}\un{cd}}F^{\ov{b}}\, \nn \\
& & - \frac{1}{2}D_{\ov{a}}F_{\ov{b}\un{cd}}F^{\ov{a}\un{cd}}F^{\ov{b}} - \frac{1}{4}D_{\un{a}}F_{\ov{b}\un{de}}F_{\ov{c}}{}^{\un{de}}F^{\un{a}\ov{bc}} - D_{\ov{a}}F_{\ov{b}\un{cd}}F^{\ov{a}\un{ce}}F^{\ov{b}\un{d}}{}_{\un{e}} + \frac{1}{4}D_{\ov{a}}F_{\ov{b}\un{de}}F^{\ov{c}\un{de}}F^{\ov{ab}}{}_{\ov{c}}\, \nn \\
& & + \frac{1}{4}F_{\ov{a}\un{cd}}F^{\ov{a}}{}_{\un{ef}}F^{\ov{b}\un{ce}}F_{\ov{b}}{}^{\un{df}} - \frac{1}{4}F_{\ov{a}\un{cd}}F^{\ov{a}\un{ce}}F^{\ov{b}\un{df}}F_{\ov{b}\un{ef}} + \frac{1}{4}F_{\ov{a}\un{de}}F^{\ov{b}\un{de}}F^{\ov{ac}\un{f}}F_{\ov{bc}\un{f}} + \frac{1}{3}F_{\ov{a}\un{d}}{}^{\un{e}}F_{\ov{b}}{}^{\un{df}}F_{\ov{c}\un{ef}}F^{\ov{abc}}\, , \nn
\eea
\bea
\label{ap-Rplus}
{\cal R}^{(+)} & = & - \frac{1}{4}D_{\un{a}}D^{\un{b}}F^{\un{a}\ov{cd}}F_{\un{b}\ov{cd}} - \frac{1}{4}D_{\un{a}}D^{\un{b}}F_{\un{b}\ov{cd}}F^{\un{a}\ov{cd}} + \frac{1}{8}D_{\ov{a}}F_{\un{b}\ov{cd}}D^{\ov{a}}F^{\un{b}\ov{cd}} - \frac{1}{4}D_{\un{a}}F^{\un{a}\ov{cd}}D^{\un{b}}F_{\un{b}\ov{cd}}\, \\
& & - \frac{1}{8}D_{\un{a}}F_{\un{b}\ov{cd}}D^{\un{a}}F^{\un{b}\ov{cd}} - \frac{1}{2}D_{\un{a}}F^{\un{b}}F^{\un{a}\ov{cd}}F_{\un{b}\ov{cd}} - \frac{1}{4}F_{\un{a}}F^{\un{b}}F^{\un{a}\ov{cd}}F_{\un{b}\ov{cd}} - \frac{1}{2}D_{\un{a}}F^{\un{a}\ov{cd}}F_{\un{b}\ov{cd}}F^{\un{b}}\, \nn \\
& & - \frac{1}{2}D_{\un{a}}F_{\un{b}\ov{cd}}F^{\un{a}\ov{cd}}F^{\un{b}} - \frac{1}{4}D_{\ov{a}}F_{\un{b}\ov{de}}F_{\un{c}}{}^{\ov{de}}F^{\ov{a}\un{bc}} - D_{\un{a}}F_{\un{b}\ov{cd}}F^{\un{a}\ov{ce}}F^{\un{b}\ov{d}}{}_{\ov{e}} + \frac{1}{4}D_{\un{a}}F_{\un{b}\ov{de}}F^{\un{c}\ov{de}}F^{\un{ab}}{}_{\un{c}}\, \nn \\
& & + \frac{1}{4}F_{\un{a}\ov{cd}}F^{\un{a}}{}_{\ov{ef}}F^{\un{b}\ov{ce}}F_{\un{b}}{}^{\ov{df}} - \frac{1}{4}F_{\un{a}\ov{cd}}F^{\un{a}\ov{ce}}F^{\un{b}\ov{df}}F_{\un{b}\ov{ef}} + \frac{1}{4}F_{\un{a}\ov{de}}F^{\un{b}\ov{de}}F^{\un{ac}\ov{f}}F_{\un{bc}\ov{f}} + \frac{1}{3}F_{\un{a}\ov{d}}{}^{\ov{e}}F_{\un{b}}{}^{\ov{df}}F_{\un{c}\ov{ef}}F^{\un{abc}}\, , \nn
\eea
with the two contributions related by the exchange of projectors, ${\cal R}^{(+)} = {\cal R}^{(-)}\left[P \leftrightarrow \bar{P}\right]$.

Different choices of $(a,b)$ then reproduce distinct theories. Specifically, $(\alpha',0)$ yields the heterotic string, $(\alpha',\alpha')$ the bosonic string, and $(\alpha',-\alpha')$ the HSZ theory. In the latter case, Riemann-squared terms are absent, and the first-order corrections reduce solely to Chern–Simons modifications of the two-form curvature.

\bibliographystyle{JHEP}
\bibliography{main}

\end{document}